# The challenges of measuring spin Seebeck noise

Renjie Luo[1], Xuanhan Zhao[1], Tanner J. Legvold[1], Liyang Chen[2], Changjiang Liu[3], Deshun Hong[3], Anand Bhattacharya[3], Douglas Natelson[1,4,5]

[1]Department of Physics and Astronomy, Rice University, Houston, TX 77005, USA
[2]Applied Physics Program, Smalley-Curl Institute, Rice University, Houston, TX 77005, USA
[3]Materials Science Division, Argonne National Laboratory, Lemont, IL 60439, USA
[4]Department of Electrical and Computer Engineering, Rice University, Houston, TX 77005, USA
[5]Department of Materials Science and NanoEngineering, Rice University, Houston, TX 77005, USA

**Abstract**:

Just as electronic shot noise in driven conductors results from the granularity of charge and the statistical variation in the arrival times of charge carriers, there are predictions for fundamental noise in magnon currents due to angular momentum being carried by discrete excitations. The inverse spin Hall effect as a transduction mechanism to convert spin current into charge current raises the prospect of experimental investigations of such magnon shot noise. Spin Seebeck effect measurements have demonstrated the electrical detection of thermally driven magnon currents and have been suggested as an avenue for accessing spin current fluctuations. Using spin Seebeck structures made from yttrium iron garnet on gadolinium gallium garnet, we demonstrate the technical challenges inherent in such noise measurements. While there is a small increase in voltage noise in the inverse spin Hall detector at low temperatures associated with adding a magnetic field, the dependence on field orientation implies that this is not due to magnon shot noise. We describe theoretical predictions for the expected magnitude of magnon shot noise, highlighting ambiguities that exist. Further, we show that magnon shot noise detection through the standard inverse spin Hall approach is likely impossible due to geometric factors. Implications for future attempts to measure magnon shot noise are discussed.

I. Introduction

With the advent of the spin Hall and inverse spin Hall effects (SHE, ISHE), angular momentum transport phenomena in magnetic insulators have become measurable electrically, providing a means of examining emergent spin-carrying excitations. In a magnetic insulator with magnons, angular momentum transport is expected to exhibit a "spin shot noise", since each discrete magnon carries an angular momentum of magnitude $\hbar$. [1] Charge shot noise, the intrinsic fluctuations in the electrical current resulting from the arrival of discrete electronic charges, was predicted more than a century ago [2] and has been an invaluable tool for examining the nature and statistics of current-carrying excitations, including systems with unusual emergent charge degrees of freedom. [3-8] A fluctuating magnon current is expected to lead to corresponding fluctuations in the inverse spin Hall voltage in a properly oriented detector. [9] Spin shot noise, like charge shot noise [10], is expected to be white noise (independent of frequency) at low frequencies and would produce an additional spin-based contribution to the voltage fluctuations on the ISH detector. Spin shot noise situations examined theoretically include coherently driven and thermally driven magnon currents, [11] magnon squeezing, [12,13] and suppression of the spin shot noise in the diffusive limit. [14] To date there are no reported experimental measurements of spin (magnon) shot noise in magnetic insulators. Spin shot noise has been highlighted as a potential probe of emergent spin-carrying excitations in spin liquids, [15,16] spin ice, [17,18] and other systems with unusual order. [19,20]

One way to drive a magnon current in a magnetic insulator is through thermal excitation via the spin Seebeck effect (SSE). In analogy with the conventional Seebeck effect, an applied temperature gradient $\nabla T$ generates an angular momentum current carried by a net flux of magnons from hot to cold. Interfacial exchange processes between the magnons in the magnetic insulator and the conduction electrons in a strong spin-orbit (SO) metal wire can transfer spin current into the metal. For properly oriented magnetization, that spin current generates a charge current in the strong SO wire through the ISHE, leading to a spin Seebeck voltage. [21] Previously, insulating ferrimagnets (FMIs) such as $Y_3Fe_5O_{12}$ (yttrium iron garnet, YIG) have shown SSE response in the local configuration, [22-26] and the local SSE has also been seen [27] in the insulating paramagnet (PM), $Gd_3Ga_5O_{12}$ (GGG), which is a geometrically frustrated magnetic material. Recent work [28] shows that the main SSE contribution in many experiments originates from the temperature gradient across the bulk magnetic insulator, and that phonon drag [29] can be a dominant factor in driving magnon currents. In general, however, systematic experiments in different sample geometries are required to delineate between bulk and interfacial contributions to the SSE response [30]. For the purposes of the present work, the only essential point is that some spin current (originating from discrete magnons) be transmitted into the ISH detector.

As mentioned, spin (magnon) shot noise – fluctuations in the magnon current (white noise) due to the discreteness of magnon excitations – should produce resultant fluctuations (white noise) in the ISH voltage. At a constant temperature, any ISH detector will also exhibit intrinsic voltage fluctuations in the absence of any magnon or thermal current due to Johnson-Nyquist noise, the thermal charge noise due to the fluctuation-dissipation theorem [31,32]. In this work we use local SSE devices based on YIG films on GGG substrates, a well-studied system, to demonstrate the technical challenges of this spin shot noise measurement approach, and we argue that measuring spin shot noise using standard ISH geometries is likely not possible. As expected, the ISHE spin Seebeck voltage increases linearly with applied heater power, consistent with the effective spin current transferred to the Pt being proportional to the heater power and hence local temperature gradients. The voltage noise power, $S_V$, in the Pt increases with heater power as well, even in zero magnetic field, indicating warming of the Pt. A small additional contribution to the noise is present when a large in-plane magnetic field is applied, but this lacks the field orientation dependence expected for spin shot noise. We discuss the challenges in interpreting this data and argue that successful and conclusive measurements of the spin shot noise will require advances in both the theoretical understanding and modeling of this problem. An analysis of the sample geometry dependence of ISHE detection of spin currents shows that changes in experimental geometry and technique are essential to any eventual detection of spin shot noise.

## II.    Experimental Configuration and Procedure

Figure 1 shows a schematic of the typical device geometry examined in this work, very similar to that used in other studies of the local SSE [23,27,33,34]. The purpose of these experiments is to show the experimental challenges inherent in magnon/spin shot noise measurements in this SSE/ISH configuration. Devices measured include structures fabricated at Argonne National Laboratory as well as those fabricated at Rice University. All devices were based on a YIG film of 84 nm thickness deposited by magnetron sputtering on GGG (111) substrates at Argonne. Film quality is high (see supporting material for x-ray reflectometry characterization). In Argonne-made SSE devices, the wires were patterned by photolithography and Ar ion milling, with a device stack consisting (bottom-to-top) of an 800 μm by 10 μm sputtered Pt stripe 5 nm thick, a 100 nm $Si_3N_4$ barrier for electrical isolation, and a 50 nm thick, 10 μm width Au heater wire with a resistance of approximately 100 Ω. In Rice-made SSE devices, the wires were patterned through photolithography and liftoff processing. The device stack is similarly (bottom-to-top) a sputtered Pt inverse spin Hall detector 7.5 nm thick, 10 μm wide, and 800 μm long, a 100 nm thick evaporated $SiO_x$ barrier for electrical isolation, a 5 nm thick, 12 μm wide evaporated Ti layer (acting as adhesion for the Au), and a 50 nm thick, 12 μm wide, 700 μm long, e-beam evaporated Au heater wire with resistance of approximately 60 Ω at room temperature.

Devices are mounted on a custom shielded probe in a Quantum Design Physical Property Measurement System. For a given device, the resistance as a function of temperature is measured for the Au heater ($R_{Au}$) and Pt detector ($R_{Pt}$) down to low temperatures (typically 5 K). Similarly, at low temperatures the resistances of both heater and detector are measured as a function of magnetic field oriented in the plane of the sample. To measure spin Seebeck response, the heater is driven with an AC current at angular frequency $\omega$ (= $2\pi \times 7.7$ Hz) and the ISH voltage on the Pt wire segment is detected at $2\omega$ with a lock-in amplifier. The dominant temperature gradient in this configuration is expected to be along the $z$ direction as indicated in Fig. 1b. While one must be careful about the possibility of in-plane temperature gradients having some consequences, finite-element thermal modeling (see Supplemental Material [35] Fig. S8, S9 and accompanying text, as well as Refs. [36,37] therein) confirms that the largest temperature gradient is along $z$. This AC demodulation technique to measure the SSE response gives greatly improved signal to noise compared to applying a DC current to the heater, avoiding confounding effects from extrinsic noise pickup.

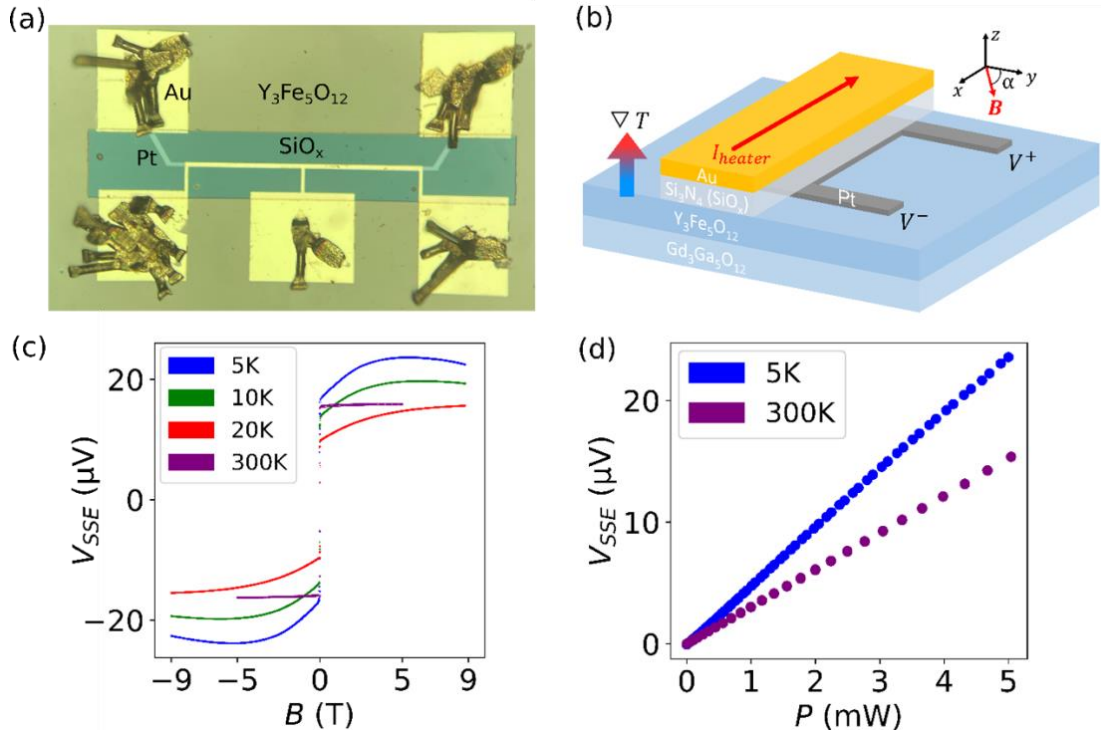

FIG 1. (a) Optical micrograph of a Rice-made device. (b) Illustration of device and measurement configuration. (c) Field dependence of the SSE on the Rice-made 10 μm wide device at selected temperatures for a heater power of 5 mW, with field in-plane and transverse to the Pt wire; Argonne-made devices show quantitatively similar responses (see Supplemental Material Fig. S4). (d) Dependence of $V_{SSE}$ on heater power at fixed cryostat temperature and 0.5 T field applied in-plane, transverse to the Pt wire.

Figure 1c shows the SSE response at a fixed heater power of 5 mW as a function of magnetic field applied transverse to the Pt wire. In this field orientation, the SSE voltage has two contributions: A component from the magnetization of the YIG that switches near zero field, and a component at high fields that, based on its temperature and field dependence, originates with the paramagnetic response of the GGG substrate. The precise nature of spin transport through GGG remains a topic of investigation [27,38,39].

Note that a SSE response related to the GGG's magnetization implies that the temperature gradient across the different components of the whole stack is important, but not necessarily that there is bulk spin current generated in the GGG. The coupling of spin current in the magnetic insulator to the conduction electrons of the Pt detector takes place at the YIG/Pt interface. One possible explanation for the appearance of a GGG contribution to the signal could be a bulk magnon current in the GGG itself [28] impinging on the YIG. It is also possible, however, that a temperature difference across the YIG/GGG interface could generate spin current and a magnon chemical potential in the YIG without bulk spin current in the GGG [40]. It is challenging in general to assess experimentally the relative importance of bulk-generated spin currents vs. interfacially-generated spin currents [30], as assessing interfacial temperature differences across metal/insulator and insulator/insulator boundaries is difficult. This issue, whether the SSE-related spin current originates from bulk temperature gradients across the magnetic insulator, interfacial temperature differences between magnetic insulators, or the local temperature difference, $T_{Pt} - T_{YIG}$, is an important distinction when attempting to estimate the expected magnitude of a spin shot noise signature, as explained further below. In principle, regardless of bulk vs. interfacial origins of the spin current in the stack, a spin shot noise signature is expected when the spin transfer to the ISH detector originates from discrete spin-carrying magnon excitations.

The SSE voltage varies linearly in the heater power (Fig. 1d), and the field-dependence and magnitude of the SSE voltage is comparable to what has been seen in previous experiments on these materials. A response is seen in zero applied field because of the previously coerced magnetization of the YIG. For the case when the magnetic field is oriented in the plane of the sample, parallel to the Pt wire, the magnitude of the second-harmonic voltage is much lower and shows a much weaker field dependence (see Supplemental Material [35] Fig. S1). This is consistent with expectations, since for a perfectly parallel orientation of the magnetization in the sample, a thermally driven magnon flux would not be transporting angular momentum with the proper orientation to generate an ISH voltage along the Pt wire. Complete dependence of the SSE voltage as a function of the direction of the field in the sample plane for the 10 μm width Rice-made device obtained using a sample-rotation probe is shown in Fig. S7.

As shown in Fig. 2, when measuring voltage noise in the Pt wire, heater power is applied via a DC current in the Au heater wire, filtered to minimize extrinsic noise pickup. The voltage noise across the Pt wire is acquired through a standard cross-correlation method [41], using two voltage amplifier chains in parallel. In each amplifier chain, the first-stage voltage preamplifier (LI-75A, gain = 100) is AC-coupled. The second-stage amplifier (SR560, gain = 100) is fed into a high-speed digitizer (Picoscope 4262). The digitizer inputs are cross-correlated to minimize the impact of amplifier input noise. The digitizer samples at 5 MHz with 10 ms for each time series, and a final spectrum is an average of 300 of such series. The raw noise spectrum rolls off at high frequencies due to capacitance in the wiring and the resistance of the Pt wire. The gain of the noise system is calibrated in two ways, with a series of known resistors at fixed temperature and with a single resistor near the Pt resistance value as a function of temperature, comparing the measured spectra to the calculated Johnson-Nyquist (JN) voltage noise power per unit bandwidth, $S_{V,JN} = 4k_B TR$, where $R$ is the relevant resistance.

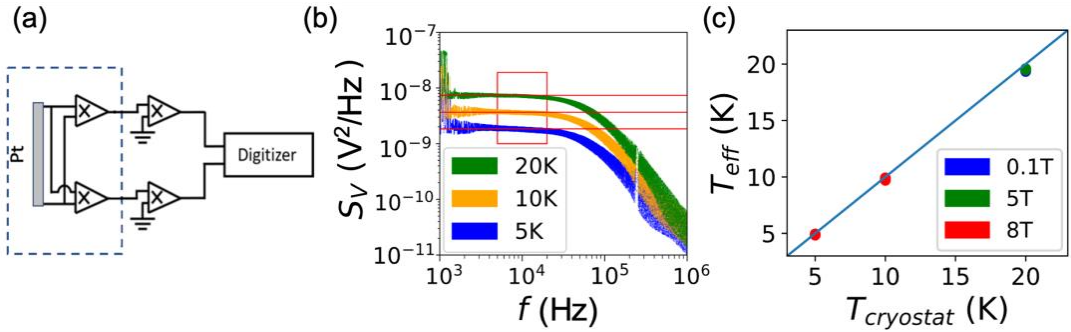

FIG 2.: (a) Schematic of the noise measurement setup. (b) Example spectra at various temperatures from a Rice-made 10 μm wide device. The roll-off at high frequencies is due to capacitive effects of the wiring combined with the Pt wire ISH detector resistance. Large extrinsic noise peaks have been filtered out by removing the top and bottom 20% of data points per unit frequency on a log scale. The median, giving $S_{V,0}$, is computed from the data in the red box and displayed as a red line for each spectrum. (c) Measured effective temperature of the Pt wire vs. reported temperature of the cryostat at zero heater power, field oriented in the plane transverse to the Pt wire. All points are shown and overlap.

It is convenient and intuitive to characterize the magnitude of the noise in the Pt ISH detector wire, $S_{V,0}$, in terms of an effective temperature, $T_{eff} = \frac{S_{V,0}}{4k_B R_{Pt}}$, based on the measured Pt resistance, $R_{Pt}$. In the absence of any magnetic contribution to the noise, $T_{eff}$, is purely from the JN noise and is a measure of the local temperature of the Pt wire, which will increase above the cryostat temperature when the heater is driven. Since the JN noise is unavoidably present, $T_{eff}$ is a simple way to think about the relative magnitude of any changes in the noise due to, e.g., a desired spin-dependent contribution, or an undesired temperature drift. These noise measurements are performed at constant DC heater power rather than with an AC modulation of the heater. (This avoids any capacitive pickup of a strong narrow-band feature in the noise spectrum at the would-be AC drive frequency and allows the necessary averaging to acquire clean noise spectra.) At a given field, heater power, temperature or device orientation, 10 spectra (which are themselves averages of 300 individual spectra) are consecutively recorded. The spectra obey $S_V(f) = \frac{S_{V,0}}{1+(R_{Pt}C)^2 f^2}$, a source Johnson-Nyquist noise density, $S_{V,0}$, low-pass filtered by the capacitance, $C$, of the wiring. The $S_{V,0}$ of a spectrum could be determined as a nonlinear fit parameter but is obtained more robustly as the median of all data points in the range 5 kHz to 20 kHz since extrinsic noise peaks can give large residuals, interfering with averages and fits. The range 5 kHz to 20 kHz was chosen because it is a relatively "quiet" region across all trials that is negligibly low-pass filtered by capacitance. We then calculate the mean and standard error of these 10 effective temperatures to produce the individual data points of Fig. 3.

With the measured Pt resistance, the JN noise at zero applied field and zero applied heater power provides an accurate thermometer of the sample temperature. For a given device, at each field orientation, and heater power, the probe is allowed to equilibrate for 60 minutes after the cryostat temperature has stabilized, to avoid temperature drifts. Sample temperature stability data from JN noise is shown in Fig. S10. The noise spectrum in the Pt is measured repeatedly at multiple heater powers, $\dot{q}$, both at a low field (zero field, or 0.1 T to have a definitive coerced direction for the YIG magnetization) and at high fields (5 T, 8 T). The averaged noise spectra between 5 kHz and 20 kHz are used to characterize the noise response of the Pt wire in terms of an effective temperature increase from the zero-heater-power case: $\Delta T_{eff} = (S_V(\dot{q}, B) - S_V(\dot{q}=0, B))/4k_B R_{Pt}(B)$.

### III. Experimental Results and Discussion

The change in voltage noise with heater power is clear at a cryostat temperature of 5 K. As is shown in Fig. 3a, for magnetic field transverse to the 10 μm wide Pt wire (green in Fig. 3), there is a resolvable effective temperature increase that is larger for high magnetic fields. For a heater power of 5 mW and a cryostat temperature of 5 K, $\Delta T_{eff} \approx 2.25$ K at 0.1 T and $\Delta T_{eff} \approx 2.6$ K at 8 T. Already it is clear that any magnetic field-dependent contribution to the noise is small compared to both the change in JN noise just from having the heater on, and the baseline JN noise of the Pt ISH detector at zero heater power. Resolving such a small difference in noise spectra requires excellent temperature stability of the measurement system and can easily be hampered by drifts; hence the 60 minute stabilization time prior to data acquisition, which reduces temperature drifts over the timescale of noise spectra acquisition to ∼ 10 mK. Similarly, such small changes in the noise require careful measurements of the magnetoresistance of the Pt ISH detector, since changes as a function of field $R(B) - R(B = 0)$ lead directly to changes in the JN noise of the Pt wire that must be considered.

Measurements at higher cryostat temperatures struggle to resolve field dependence in $\Delta T_{eff}$, as shown in Fig. 3b,c. There are two reasons for this. First, at 10 K and 20 K, the $\Delta T_{eff} \approx$ 1.01 K and 0.63 K at 0.1 T, respectively, for the same 5 mW heater power, so that any expected field dependence would be smaller relative to that seen at the lower cryostat temperature. Second, the measurement-to-measurement variation in the spectra that give $\Delta T_{eff}$ itself is larger at higher temperatures.

The decreased magnitude of any effect at elevated temperatures is expected on general grounds. As $T$ is increased, the thermal conductivity of all the constituent materials increases, so that at fixed power the temperature gradients across all the materials are proportionately smaller. Similarly, thermal boundary resistances [42] at all metal-dielectric interfaces become less important at elevated temperatures. An analogous decrease in $\Delta T_{eff}$ at fixed heater power with increasing cryostat temperature has been observed previously in JN noise measurements used to characterize the magnitude of the Nernst-Ettingshausen response in Pt thin films [43].

An important control experiment is the measurement of the ISH detector voltage noise under heater power when the magnetic field is oriented *parallel* to the Pt wire. In that configuration, $V_{ISH}$ from any magnon flux should ideally be zero if the field alignment is perfect and there is no residual misaligned magnetization from the YIG. As shown in Figure 3, there is still a non-zero change in noise with field in the parallel field orientation, comparable to that seen in the perpendicular field orientation.

This suggests that the field-dependent change in noise when heater power is applied likely has its origins in the field-dependent thermal conductivity of the material stack rather than the spin shot noise. This is surprising, given that measurements and thermal modeling in similar geometries show that the dominant thermal resistances in such structures tend to be at metal/dielectric boundaries [43], and that the low temperature thermal conductivity of GGG can be modeled well as being phonon-based, with little magnetic field dependence [44]. Another concern could be that the thermometry of the measurement system is affected noticeably by magnetic field (that is, the cryostat temperature feedback is not properly accounting for the magnetoresistance of its thermometers), but Quantum Design calibrations already account for this to better than $\frac{\Delta T}{T} < 0.01$, and no such field dependence is observed in Pt devices without magnetic materials [43]. With the small sizes of signals expected in spin shot noise measurements, it is clear that a detailed understanding of the thermal path and thermometry is essential.

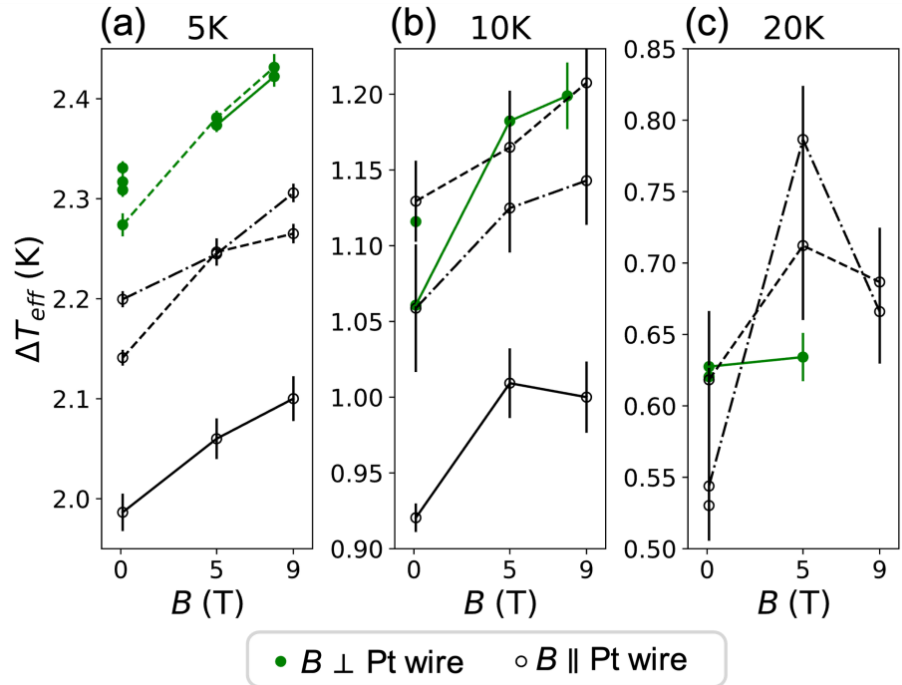

FIG 3. Average effective temperature change for various cryostat temperatures, applied magnetic fields (0.1 T, 5 T, 8 T, and 9 T), and orientations (filled green is field transverse to the

Pt wire, empty black is field parallel to the Pt wire) for the 10 μm wide Rice device. Each data point is a mean of the effective temperature changes of 10 spectra (which are themselves averages of 300 individual spectra); the bars are the standard error of the means. Data points connected by lines were taken in the same "run" and thus should have similar baselines even in the presence of slow thermal drift; differently styled lines indicate different runs (sequences of field at each temperature).

Summarizing the experimental situation and the challenges of such measurements: Any contribution to the voltage noise in a Pt ISH detector due to shot noise in the magnon current would add directly to the JN noise already present in the detector. This implies that fine temperature stability is required for any such measurements to be plausible, to avoid confounding noise changes from temperature drifts. The heater which drives the SSE also elevates the ISH detector's temperature, leading to an increase in the JN noise, so any spin contribution would have to be detected through the magnetic field magnitude and orientation dependence of any heater-driven noise. Measurements in Pt on YIG/GGG devices show detectable excess noise at high magnetic fields, but this noise does not depend on the orientation of the magnetic field, in contradiction to expectations for a true magnon shot noise contribution. This suggests that the field dependence of the noise results from field-dependent changes in the Pt ISH detector temperature due to field-dependent thermal conductance of the device stack. These measurements highlight key experimental challenges inherent in attempts to measure SSE-based spin shot noise.

## IV. Modeling and Analysis

When interpreting the data above, it is important to consider the expected magnitude of any noise contribution due to fluctuations in the magnon current. As explained below, there are ambiguities in the modeling of the processes that lead to the spin shot noise in this spin Seebeck scenario. Beyond these resolving these issues, we find that geometric factors make the detection of spin shot noise via the ISHE likely impossible.

First we consider the relationship between the measured spin Seebeck voltage and the spin current into the Pt. For an inverse spin Hall detector [39] with spin Hall angle $\theta_{SH}$, electrical conductivity $\sigma_N$, thickness $t_N$, and spin diffusion length $\lambda_N$, when an appropriately oriented spin current density $j_s$ is transmitted across the magnetic insulator/spin Hall metal interface, the induced inverse spin Hall electric field in the detector is $E_{ISH} = \left(\frac{\theta_{SH}}{\sigma_N}\right)\left(\frac{\lambda_N}{t_N}\right)\left(\frac{2e}{\hbar}\right) j_s \tanh\left(\frac{t_N}{2\lambda_N}\right)$. The terms involving the ratio of $t_N/\lambda_N$ account for the decay of the spin current as it penetrates into the nonmagnetic Pt layer. For a detector of width $w_N$ and length $L$, the net transmitted spin current (of proper spin orientation) is $I_s = j_s \times L w_N$, giving the ISH voltage:

$$V_{ISH} = \left(\frac{\theta_{SH}}{\sigma_N}\right)\left(\frac{\lambda_N}{t_N}\right)\left(\frac{2e}{\hbar}\right)\left(\frac{I_s}{w_N}\right) \tanh\left(\frac{t_N}{2\lambda_N}\right) \tag{1}$$

and therefore $I_s$ into the ISH detector in terms of the measured ISH voltage:

$$I_s = V_{ISH} w_N \left(\frac{\sigma_N}{\theta_{SH}}\right)\left(\frac{t_N}{\lambda_N}\right)\left(\frac{\hbar}{2e}\right) \coth\left(\frac{t_N}{2\lambda_N}\right). \tag{2}$$

Note that the total magnon spin current within the magnetic insulator is larger than this because there is a spin current resistance associated with the conversion efficiency of the YIG/Pt interface. [45]

There is an argument in the literature [11,46] that the proper relationship between the spin-induced charge current noise $S_{I,ISH}$ and the spin current noise $S_s$ should be (ignoring the decay of spin current into the Pt)

$$S_{I,ISH} = \left(\frac{2e}{\hbar}\right)^2 \theta_{SH}^2 S_s, \tag{3}$$

and therefore the charge voltage noise $S_{V,ISH}$ should be

$$S_{V,ISH} = R_{Pt}^2 \left(\frac{2e}{\hbar}\right)^2 \theta_{SH}^2 S_s. \tag{4}$$

Rewriting in terms of $R_{Pt} = L/(\sigma_N t_n w_n)$, this gives

$$S_{V,ISH} = \left(\frac{\theta_{SH}}{\sigma_N}\right)^2 \left(\frac{L}{t_N}\right)^2 \left(\frac{2e}{\hbar}\right)^2 \left(\frac{1}{w_N}\right)^2 S_s. \tag{5}$$

This result shows a scaling with device length, for a given spin shot noise, that suggests that longer devices would favor detectability of the spin shot noise contribution to the voltage fluctuations.

However, Eq. (4) is too simplistic a treatment of the link between the ISH voltage noise and the spin current noise in a real device geometry. Eq. (1) treats the geometric factors appropriately in an ISHE detection configuration, giving a relationship between $V_{ISH}$ and $I_s$. From that equality, one should consider mean square fluctuations per unit bandwidth of the quantities on both sides, which would imply:

$$S_{V,ISH} = \left(\frac{\theta_{SH}}{\sigma_N}\right)^2 \left(\frac{\lambda_N}{t_N}\right)^2 \left(\frac{2e}{\hbar}\right)^2 \left(\frac{1}{w_N}\right)^2 \tanh^2\left(\frac{t_N}{2\lambda_N}\right) S_S \quad , \tag{6}$$

This contradicts Eq. (4), and Eq. (6) predicts a value for the spin shot noise contribution to the Pt voltage noise that is smaller than that of Eq. (4) (assuming the tanh² factor is of order 1) by a factor of $\left(\frac{\lambda_N}{L}\right)^2$, which can easily be on the order of $10^{-10}$. This strongly implies that spin shot noise is not a measurable quantity using ISH detection in the geometry typically used for SSE or spin transport measurements. This is a key take-away from this work.

Additionally, the dependence of the expected spin current noise $S_s$ on the magnitude of the spin current $I_s$ is also a nontrivial issue, as the system must be driven far enough from equilibrium to generate magnon shot noise. In equilibrium ($I_s = 0$), in accordance with the fluctuation-dissipation theorem, any noise involving thermally distributed magnons interacting with the Pt conduction electrons should already be captured by the Johnson-Nyquist noise through the spin Hall contribution to the Pt resistance. Exactly this equilibrium noise contribution has been observed in Pt on YIG. [47] In the zero-temperature, nonequilibrium limit of a Poissonian arrival of uncorrelated spin-1 magnons, the expected form of the spin current noise would be $S_s = 2\hbar I_s$, in analogy with the charge current shot noise for Poissonian charges of magnitude $e$, $S_I = 2eI$. Deviations from the Poisson arrival statistics are conventionally wrapped into a multiplicative Fano factor, $F$, such that $S_I = F \cdot 2eI$. The Fano factor captures whether the shot noise is sub-Poissonian (carriers are "antibunched") or super-Poissonian (carriers are "bunched") relative to the simple Poisson statistics. There are predictions in the literature for super-Poissonian spin shot noise [12] in some magnon systems, which would in principle make spin shot noise easier to detect.

The precise form of the expected crossover from equilibrium spin current noise to nonequilibrium spin shot noise as a function of driving is important to understand. In the case of a device such as a tunnel junction that exhibits *charge* shot noise ($F = 1$), the finite temperature current noise is given by $S_I = 2eI \coth\left(\frac{eV}{2k_BT}\right)$, where $I$ is the current and $V$ is the applied bias. This reduces to the Johnson-Nyquist result $S_I = 4k_BT\left(\frac{I}{V}\right)$ in the limit of zero bias (equilibrium), and approaches $2eI$ in the high bias limit when the system is driven sufficiently far from equilibrium. In the charge shot noise case, voltage bias plays a double role, driving the system out of equilibrium to generate the current and determining the energetic range available for electronic transport because of the electronic Fermi distribution. In the spin Seebeck effect case, the magnons are bosonic. The question is, if one wants to observe spin shot noise in the something like the longitudinal SSE geometry, what is the appropriate measure of how far the system is driven from equilibrium? A closely related question is, should the spin shot noise depend on the frequency of the magnons involved? This is an issue examined in Ref. [9] for the spin pumping case and Ref. [11] for the spin Seebeck case.

Matsuo *et al.* [11,48] consider angular momentum shot noise in the spin Seebeck case. They argue that the relevant criterion to observe spin shot noise in thermally driven magnon transport is that the driving temperature difference be $\frac{|T_h - T_c|}{T} \gg 2k_BT/g\mu_BB$ at low temperatures, $\frac{k_BT}{g\mu_BB} \ll 1$, where $T_h - T_c$ is the temperature difference between the hot side ($T_h$) and the cold side ($T_c$) and $T$ is the average temperature. In this model $g\mu_BB = \hbar\omega_0(B)$ is the field-induced gap in the spin-wave spectrum of the YIG. The resulting predicted spin shot noise does not directly depend on the frequency of the excited magnons. Note that these authors consider the SSE as resulting only from the temperature difference between the magnetic insulator and the SO coupled metal, while in the present experiment the SSE voltage clearly involves the temperature gradient across the bulk of the GGG/YIG stack. Still, in this approach, the expected form of the spin shot noise would be

$$S_s = 2\hbar I_s \coth\left(\frac{\hbar\omega_0(B)}{2k_B}\left(\frac{T_h - T_c}{T_h T_c}\right)\right) \qquad (7)$$

One can take the Laurent expansion for $\coth(x) = \left(\frac{1}{x}\right) + \left(\frac{x}{3}\right) - \left(\frac{x^3}{45}\right) + \cdots$ and keep only the first two terms. The average spin current $I_s \propto (T_h - T_c)$, so the first term in the expansion will lead to a Johnson-Nyquist-like contribution independent of the driving temperature gradient (and hence heater power). The first-order term would then lead to a predicted Fano factor for the spin shot noise of $F \approx \frac{\hbar \omega_0}{6 k_B}\left(\frac{T_h - T_c}{T_h T_c}\right)$. For YIG, the spin-wave gap has been measured [49] to be $\frac{\hbar \omega_0}{k_B} \approx$ 11.2 K at 8 T.

There can be an additional factor that suppresses the noise. In charge shot noise in a diffusive mesoscopic conductor, it has been established through multiple theoretical techniques [50-53] and confirmed experimentally [54] that $F = 1/3$ for the case of weak electron-electron scattering and no electron-phonon scattering. An analogous additional Fano factor of $F = 1/3$ has been argued for diffusive magnon transport [14].

An alternative energy scale to consider is the one inferred from the spin chemical potential accumulated due to the temperature-driven magnon flow. For an ISH detector [55], the spin chemical potential in the Pt at the YIG/Pt interface is estimated to be $\mu_s = \frac{2 t_N}{\theta_{SH} L} \frac{1 + \exp\left(-\frac{2 t_N}{\lambda_N}\right)}{\left(1 - \exp\left(-\frac{t_N}{\lambda_N}\right)\right)^2} V_{ISH}$. For the situation here, this would be lower than the actual magnon chemical potential, $\mu_m$, in the YIG due to the finite interfacial spin conductance. For the numerical values assumed above, this leads to $\mu_s \approx 1.2 \times 10^{-9}$ eV for a $V_{ISH}$ of around 2 µV as seen in the data on the Argonne-made device. At 5 K, this is far smaller than $k_B T$ ($4.3 \times 10^{-4}$ eV), suggesting that it would be extremely difficult to see nonequilibrium magnon shot noise at all, if spin chemical potential in the Pt is the sole driving energy scale. The interfacial spin conductance between the magnetic insulator and the Pt ISH detector will also hamper attempts to measure the magnon shot noise. The opacity of that interface for spin means that the spin current into the Pt is reduced from the actual magnon-carried spin current in the YIG.

For a general spin shot noise Fano factor $F$, and using Eq. (2) as the relationship between $I_s$ and the measured $V_{ISH}$, the expected voltage noise in the Pt based on (the overly simplistic) Eqs. (3-5) is:

$$S_{V,ISH} = F \cdot 2\hbar \left(\frac{\theta_{SH}}{\sigma_N}\right)\left(\frac{L^2}{t_N \lambda_N}\right)\left(\frac{2e}{\hbar}\right)\left(\frac{1}{w_N}\right) \coth\left(\frac{t_N}{2\lambda_N}\right) V_{ISH} \qquad (8)$$

and based on Eq. (2) and the formulation that accounts for sample geometry (Eq. (6)) is

$$S_{V,ISH} = F \cdot 2\hbar \left(\frac{\theta_{SH}}{\sigma_N}\right)\left(\frac{\lambda_N}{t_N}\right)\left(\frac{2e}{\hbar}\right)\left(\frac{1}{w_N}\right) \tanh\left(\frac{t_N}{2\lambda_N}\right) V_{ISH} \qquad (9)$$

Typical parameters for the device made at Argonne is $t_N$ = 5 nm, $w_N$ = 10 μm, $L$ = 160 μm, and the Pt resistance $R_{Pt}$ = 2915 Ω, implying $\sigma_N$ = 1.10 × 10$^6$ S. The change in $V_{ISH}$ from 0 T to 8 T at 5 K and 5 mW heater power in that device is about 2 μV (see SM Fig. 4). A typical value for the spin Hall angle for Pt is around $\theta_{SH} \approx 0.1$, and a spin diffusion length in Pt is $\lambda_N \approx 1.5$ nm [39]. For the devices made at Rice, $t_N$ = 7.5 nm, $w_N$ = 10 μm, $L$ = 800 μm, and the Pt resistance $R_{Pt}$ = 8900 Ω, implying $\sigma_N$ =1.2 × 10$^6$ S. In Rice-made devices for 5 mW heater power at 5K the change in $V_{ISH}$ from 0 T to 8 T is roughly 20 μV.

Initially assuming $F = 1$, plugging into the (not accounting for detection geometry) Eq. (8) using the Argonne (Rice) device parameters gives an estimated spin contribution to the voltage noise of $S_V \approx 4.3 \times 10^{-17}$ V²/Hz ($S_V \approx 6.2 \times 10^{-15}$ V²/Hz). When accounting for the detection geometry properly, Eq. (9) gives an estimated voltage noise of $S_V \approx 3.3 \times 10^{-27}$ V²/Hz ($S_V \approx 2.1 \times 10^{-26}$ V²/Hz). Note that the actual measured change in voltage noise with power (from 0 mW to 5 mW) changes with field (from 0.1 T to 8 T) by about $4k_B R \, \Delta T_{eff} \approx 4.8 \times 10^{-20}$ V²/Hz ($4k_B R \, \Delta T_{eff} \approx 7.4 \times 10^{-20}$ V²/Hz).

The measured noise change with field falls between the predictions for the microscopic literature expectation Eq. (8) and the formulation Eq. (9) that accounts directly for the device geometry of the ISH transduction mechanism between spin current and ISH voltage. Based on Eq. (7) and an additional 1/3 due to spin diffusion, using $T_h = 8$ K and $T_c = 6$ K, an estimated Fano factor of 0.026 would result, but this would still leave the Eq. (8) prediction an order of magnitude larger than any measured effect in this experiment. These are additional reasons to suspect that the noise increase with field detected in these measurements is *not* due to spin shot noise.

Furthermore, for reasonable material parameters, Eq. (9) implies that measuring the SSE spin shot noise is not experimentally feasible in the kind of device geometry commonly used for SSE measurements. Based on Eq. (9), taking the tanh factor as of order 1,

$$\frac{S_{V,ISH}}{S_{V,JN}} = \theta_{SH}\left(\frac{\lambda_N}{L}\right)\left(\frac{eV_{ISH}}{k_B T}\right). \qquad (10)$$

Given that typical $V_{ISH}$ are microvolts and $\lambda_N$ is the nanometer scale, the Johnson-Nyquist noise would outweigh the spin shot noise contribution to the Pt voltage noise by orders of magnitude even at dilution refrigerator temperatures. Note that this argument does not rest on the spin Seebeck approach and would also hold for driven spin pumping methods [56] of trying to observe spin shot noise. While this manuscript was being revised, others have reached this same conclusion [57].

V.    **Conclusions**

We have demonstrated that experimental attempts to measure spin shot noise in spin Seebeck devices face many challenges, including the need for exquisite temperature stability and deep understanding of any magnetic field-dependence of the thermal paths. Measurements performed in YIG/GGG devices do show a small increase in the charge voltage noise in the Pt ISH detector at magnetic fields where the paramagnetic contribution of GGG to the spin Seebeck voltage is increasing. However, the dependence of this signal on field orientation in the plane is not consistent with expectations of a spin shot noise signature. Beyond any spin shot noise contribution, charge noise measurements in the strong SO metal detector are dominated by the Johnson-Nyquist thermal noise, and therefore are sensitive to magnetoresistive effects and any magnetic field driven changes in thermal path. Outstanding sample temperature control and stability are necessary for any experimental approach to spin shot noise.

Importantly, there is some ambiguity in the literature about the expected magnitude of the measurable spin-driven component of the charge noise signal. While the simple relationship of Eqs. (2,3) leads to predictions of readily measurable signatures, analysis rooted in the device geometry, Eqs. (6,9) find a vastly less favorable expectation, with spin shot noise contributions strongly suppressed from the favorable scenario by the ratio $\left(\frac{\lambda_N}{L}\right)^2$. Experiments trying to observe a spin shot noise effect due to magnon transport in magnetic insulators would also benefit greatly by theoretical clarity about the expected form of such noise as a function of biasing away from equilibrium toward the (in the simplest case) $S_s \to 2\hbar I_s$ high "bias" limit.

Considering the analysis of Eqs. (6,9), different measurement approaches are likely to be needed to observe the fundamental noise in spin currents in insulators due to the granularity of spin-carrying excitations. In measurements of charge shot noise, techniques have been developed that span orders of magnitude in frequency, with the ultimate limit being the achievement of full counting statistics by detection of individual charge carrier arrivals [58]. An analogous approach, detecting individual magnons [59], may be needed to realize the promise of spin shot noise as a measurable quantity.

**Acknowledgments**

The authors acknowledge useful conversations with A. Kamra, T. Kato, M. Matsuo, and J. Sinova. RL, XZ, TJL, and DN acknowledge support from NSF DMR-1704264 and DMR-2102028. LC developed and refined the noise measurement hardware and analysis software with support from DOE BES award DE-FG02-06ER46337. CL, DH, and AB acknowledge support from Materials Science and Engineering Division, U.S. DOE, Office of Basic Energy Sciences, for sample preparation. Use of the Center for Nanoscale Materials at Argonne National Laboratory, a DOE Office of Science User Facility, was supported by the U.S. Department of Energy, Office of Science, Office of Basic Energy Sciences, under Contract No. DE-AC02-06CH11357.

**References**

[1] K. Nakata, Y. Ohnuma, and M. Matsuo, Magnonic noise and Wiedemann-Franz law, Phys. Rev. B **98**, 094430 (2018).

[2] W. Schottky, Regarding spontaneous current fluctuation in different electricity conductors, Annalen Der Physik **57**, 541 (1918).

[3] M. J. M. de Jong and C. W. J. Beenakker, Doubled shot noise in disordered normal-metal--superconductor junctions, Phys. Rev. B **49**, 16070 (1994).

[4] L. Saminadayar, D. C. Glattli, Y. Jin, and B. Etienne, Observation of the *e*/3 Fractionally Charged Laughlin Quasiparticle, Phys. Rev. Lett. **79**, 2526 (1997).

[5] R. dePicciotto, M. Reznikov, M. Heiblum, V. Umansky, G. Bunin, and D. Mahalu, Direct observation of a fractional charge, Nature **389**, 162 (1997).

[6]     X. Jehl, M. Sanquer, R. Calemczuk, and D. Mailly, Detection of doubled shot noise in short normal-metal/ superconductor junctions, Nature **405**, 50 (2000).

[7]     R. Cron, M. F. Goffman, D. Esteve, and C. Urbina, Multiple-Charge-Quanta Shot Noise in Superconducting Atomic Contacts, Phys. Rev. Lett. **86**, 4104 (2001).

[8]     P. Zhou, L. Chen, Y. Liu, I. Sochnikov, A. T. Bollinger, M.-G. Han, Y. Zhu, X. He, I. Božović, and D. Natelson, Electron pairing in the pseudogap state revealed by shot noise in copper oxide junctions, Nature **572**, 493 (2019).

[9]     A. Kamra and W. Belzig, Magnon-mediated spin current noise in ferromagnet/nonmagnetic conductor hybrids, Phys. Rev. B **94**, 014419 (2016).

[10]    K. Kobayashi and M. Hashisaka, Shot Noise in Mesoscopic Systems: From Single Particles to Quantum Liquids, Journal of the Physical Society of Japan **90**, 102001 (2021).

[11]    M. Matsuo, Y. Ohnuma, T. Kato, and S. Maekawa, Spin Current Noise of the Spin Seebeck Effect and Spin Pumping, Phys. Rev. Lett. **120**, 037201 (2018).

[12]    A. Kamra and W. Belzig, Super-Poissonian Shot Noise of Squeezed-Magnon Mediated Spin Transport, Phys. Rev. Lett. **116**, 146601 (2016).

[13]    A. Kamra, E. Thingstad, G. Rastelli, R. A. Duine, A. Brataas, W. Belzig, and A. Sudbø, Antiferromagnetic magnons as highly squeezed Fock states underlying quantum correlations, Phys. Rev. B **100**, 174407 (2019).

[14]    K. Nakata, Y. Ohnuma, and M. Matsuo, Universal 1/3-suppression of magnonic shot noise in diffusive insulating magnets, Phys. Rev. B **100**, 014406 (2019).

[15]    C. Broholm, R. J. Cava, S. A. Kivelson, D. G. Nocera, M. R. Norman, and T. Senthil, Quantum spin liquids, Science **367**, eaay0668 (2020).

[16]    L. Savary and L. Balents, Quantum spin liquids: a review, Rep. Prog. Phys. **80**, 016502 (2016).

[17]    S. T. Bramwell and M. J. Harris, The history of spin ice, J. Phys.: Cond. Matt. **32**, 374010 (2020).

[18]    C. Castelnovo, R. Moessner, and S. L. Sondhi, Spin Ice, Fractionalization, and Topological Order, Ann. Rev. Cond. Matt. Phys. **3**, 35 (2012).

[19]     J. A. M. Paddison, H. Jacobsen, O. A. Petrenko, M. T. Fernández-Díaz, P. P. Deen, and A. L. Goodwin, Hidden order in spin-liquid $Gd_3Ga_5O_{12}$, Science **350**, 179 (2015).

[20]     T. Kikkawa and E. Saitoh, Spin Seebeck Effect: Sensitive Probe for Elementary Excitation, Spin Correlation, Transport, Magnetic Order, and Domains in Solids, Ann. Rev. Cond. Matt. Phys. **14**, 129 (2023).

[21]     C.-F. Pai, L. Liu, Y. Li, H. W. Tseng, D. C. Ralph, and R. A. Buhrman, Spin transfer torque devices utilizing the giant spin Hall effect of tungsten, Appl. Phys. Lett. **101**, 122404 (2012).

[22]     K.-i. Uchida, H. Adachi, T. Ota, H. Nakayama, S. Maekawa, and E. Saitoh, Observation of longitudinal spin-Seebeck effect in magnetic insulators, Appl. Phys. Lett. **97**, 172505 (2010).

[23]     S. M. Wu, J. Hoffman, J. E. Pearson, and A. Bhattacharya, Unambiguous separation of the inverse spin Hall and anomalous Nernst effects within a ferromagnetic metal using the spin Seebeck effect, Appl. Phys. Lett. **105**, 092409 (2014).

[24]     R. Ramos, T. Kikkawa, K. Uchida, H. Adachi, I. Lucas, M. H. Aguirre, P. Algarabel, L. Morellón, S. Maekawa, E. Saitoh, and M. R. Ibarra, Observation of the spin Seebeck effect in epitaxial $Fe_3O_4$ thin films, Appl. Phys. Lett. **102**, 072413 (2013).

[25]     D. Meier, T. Kuschel, L. Shen, A. Gupta, T. Kikkawa, K. Uchida, E. Saitoh, J. M. Schmalhorst, and G. Reiss, Thermally driven spin and charge currents in thin $NiFe_2O_4$/Pt films, Phys. Rev. B **87**, 054421 (2013).

[26]     K. Uchida, T. Nonaka, T. Kikkawa, Y. Kajiwara, and E. Saitoh, Longitudinal spin Seebeck effect in various garnet ferrites, Phys. Rev. B **87**, 104412 (2013).

[27]     S. M. Wu, J. E. Pearson, and A. Bhattacharya, Paramagnetic Spin Seebeck Effect, Phys. Rev. Lett. **114**, 186602 (2015).

[28]     S. M. Rezende, R. L. Rodríguez-Suárez, R. O. Cunha, A. R. Rodrigues, F. L. A. Machado, G. A. Fonseca Guerra, J. C. Lopez Ortiz, and A. Azevedo, Magnon spin-current theory for the longitudinal spin-Seebeck effect, Phys. Rev. B **89**, 014416 (2014).

[29]     R. L. Rodríguez-Suárez and S. M. Rezende, Dominance of the phonon drag mechanism in the spin Seebeck effect at low temperatures, Phys. Rev. B **108**, 134407 (2023).

[30]     E.-J. Guo, J. Cramer, A. Kehlberger, C. A. Ferguson, D. A. MacLaren, G. Jakob, and M. Kläui, Influence of Thickness and Interface on the Low-Temperature Enhancement of the Spin Seebeck Effect in YIG Films, Physical Review X **6**, 031012 (2016).

[31]     J. B. Johnson, Thermal Agitation of Electricity in Conductors, Phys. Rev. **32**, 97 (1928).

[32]     H. Nyquist, Thermal Agitation of Electric Charge in Conductors, Phys. Rev. **32**, 110 (1928).

[33]     S. M. Wu, F. Y. Fradin, J. Hoffman, A. Hoffmann, and A. Bhattacharya, Spin Seebeck devices using local on-chip heating, J. Appl. Phys. **117**, 17C509 (2015).

[34]     S. M. Wu, W. Zhang, A. Kc, P. Borisov, J. E. Pearson, J. S. Jiang, D. Lederman, A. Hoffmann, and A. Bhattacharya, Antiferromagnetic Spin Seebeck Effect, Phys. Rev. Lett. **116**, 097204 (2016).

[35]     See Supplemental Material at [URL will be inserted by publisher] for additional data on Rice-made devices, noise data on the 2 μm wide device, magnetoresistances of the Au heater and Pt detector wires, thermal modeling details, and thermal stability data.

[36]     D. G. Cahill, S. K. Watson, and R. O. Pohl, Lower limit to the thermal conductivity of disordered crystals, Physical Review B **46**, 6131 (1992).

[37]     C. Euler, P. Hołuj, T. Langner, A. Kehlberger, V. I. Vasyuchka, M. Kläui, and G. Jakob, Thermal conductance of thin film YIG determined using Bayesian statistics, Physical Review B **92**, 094406 (2015).

[38]     C. Liu, S. M. Wu, J. E. Pearson, J. S. Jiang, N. d'Ambrumenil, and A. Bhattacharya, Probing short-range magnetic order in a geometrically frustrated magnet by means of the spin Seebeck effect, Phys. Rev. B **98**, 060415 (2018).

[39]     K. Oyanagi, S. Takahashi, L. J. Cornelissen, J. Shan, S. Daimon, T. Kikkawa, G. E. W. Bauer, B. J. van Wees, and E. Saitoh, Spin transport in insulators without exchange stiffness, Nat. Comm. **10**, 4740 (2019).

[40]     K. Oyanagi, S. Takahashi, T. Kikkawa, and E. Saitoh, Mechanism of paramagnetic spin Seebeck effect, Phys. Rev. B **107**, 014423 (2023).

[41]     A. Kumar, L. Saminadayar, D. C. Glattli, Y. Jin, and B. Etienne, Experimental Test of the Quantum Shot Noise Reduction Theory, Phys. Rev. Lett. **76**, 2778 (1996).

[42] E. T. Swartz and R. O. Pohl, Thermal boundary resistance, Reviews of Modern Physics **61**, 605 (1989).

[43] R. Luo, T. J. Legvold, L. Chen, and D. Natelson, Nernst–Ettingshausen effect in thin Pt and W films at low temperatures, Appl. Phys. Lett. **122**, 182405 (2023).

[44] B. Daudin, R. Lagnier, and B. Salce, Thermodynamic properties of the gadolinium gallium garnet, $Gd_3Ga_5O_{12}$, between 0.05 and 25 K, Journal of Magnetism and Magnetic Materials **27**, 315 (1982).

[45] L. Zhu, D. C. Ralph, and R. A. Buhrman, Effective Spin-Mixing Conductance of Heavy-Metal--Ferromagnet Interfaces, Phys. Rev. Lett. **123**, 057203 (2019).

[46] M. Matsuo, Y. Ohnuma, T. Kato, and S. Maekawa, Erratum: Spin Current Noise of the Spin Seebeck Effect and Spin Pumping [Phys. Rev. Lett. 120, 037201 (2018)], Phys. Rev. Lett. **127**, 119902 (2021).

[47] A. Kamra, F. P. Witek, S. Meyer, H. Huebl, S. Geprägs, R. Gross, G. E. W. Bauer, and S. T. B. Goennenwein, Spin Hall noise, Phys. Rev. B **90**, 214419 (2014).

[48] K. Nakata, Y. Ohnuma, and M. Matsuo, Asymmetric quantum shot noise in magnon transport, Phys. Rev. B **99**, 134403 (2019).

[49] H. Man, Z. Shi, G. Xu, Y. Xu, X. Chen, S. Sullivan, J. Zhou, K. Xia, J. Shi, and P. Dai, Direct observation of magnon-phonon coupling in yttrium iron garnet, Phys. Rev. B **96**, 100406 (2017).

[50] B. L. Altshuler, L. S. Levitov, and A. Y. Yakovets, Nonequilibrium noise in a mesoscopic conductor: A microscopic analysis, JETP Lett. **59**, 857 (1994).

[51] C. W. J. Beenakker and M. Büttiker, Suppression of shot noise in metallic diffusive conductors, Phys. Rev. B **46**, 1889 (1992).

[52] K. E. Nagaev, On the shot noise in dirty metal contacts, Phys. Lett. A **169**, 103 (1992).

[53] Y. V. Nazarov, Limits of universality in disordered conductors, Phys. Rev. Lett. **73**, 134 (1994).

[54] M. Henny, S. Oberholzer, C. Strunk, and C. Schönenberger, 1/3-shot-noise suppression in diffusive nanowires, Phys. Rev. B **59**, 2871 (1999).

[55]  L. J. Cornelissen, K. J. H. Peters, G. E. W. Bauer, R. A. Duine, and B. J. van Wees, Magnon spin transport driven by the magnon chemical potential in a magnetic insulator, Phys. Rev. B **94**, 014412 (2016).

[56]  A. Kamra and W. Belzig, Spin Pumping and Shot Noise in Ferrimagnets: Bridging Ferro- and Antiferromagnets, Phys. Rev. Lett. **119**, 197201 (2017).

[57]  L. Siegl, M. Lammel, A. Kamra, H. Huebl, W. Belzig, and S. T. B. Goennenwein, Electrical detectability of magnon-mediated spin current shot noise, Physical Review B **108**, 144420 (2023).

[58]  S. Gustavsson, R. Leturcq, B. Simovič, R. Schleser, T. Ihn, P. Studerus, K. Ensslin, D. C. Driscoll, and A. C. Gossard, Counting Statistics of Single Electron Transport in a Quantum Dot, Phys. Rev. Lett. **96**, 076605 (2006).

[59]  D. Lachance-Quirion, S. P. Wolski, Y. Tabuchi, S. Kono, K. Usami, and Y. Nakamura, Entanglement-based single-shot detection of a single magnon with a superconducting qubit, Science **367**, 425 (2020).

# The challenges of measuring spin Seebeck noise: Supplementary Material

Renjie Luo[1], Xuanhan Zhao[1], Tanner J. Legvold[1], Liyang Chen[2], Changjiang Liu[3], Deshun Hong[3], Anand Bhattacharya[3], Douglas Natelson[1,4,5]

[1]Department of Physics and Astronomy, Rice University, Houston, TX 77005, USA
[2]Applied Physics Program, Smalley-Curl Institute, Rice University, Houston, TX 77005, USA
[3]Materials Science Division, Argonne National Laboratory, Lemont, IL 60439, USA
[4]Department of Electrical and Computer Engineering, Rice University, Houston, TX 77005, USA
[5]Department of Materials Science and NanoEngineering, Rice University, Houston, TX 77005, USA

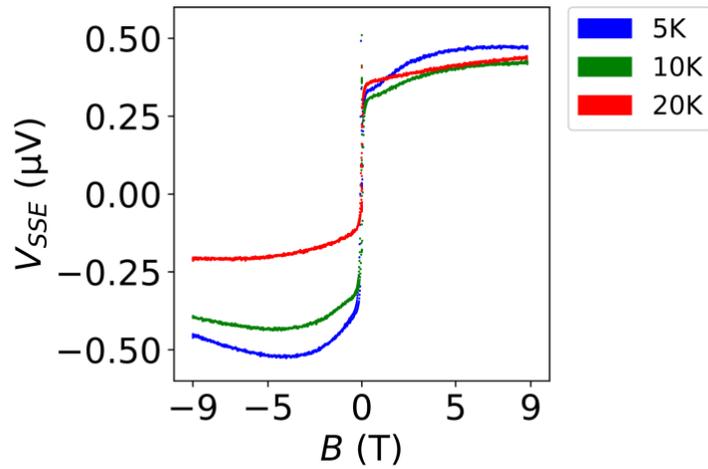

**Figure S1: SSE of the 10 μm wide Rice-made device with field parallel to the wire**

The signal is 40 times weaker than in the transverse case (Fig. 1c in the main paper) because the magnon current produced by the temperature gradient does not carry spin of the appropriate orientation to drive a inverse spin Hall voltage along the wire. The effect is still weakly seen because the field and device are slightly off true parallel. Recorded with a power of 5 mW.

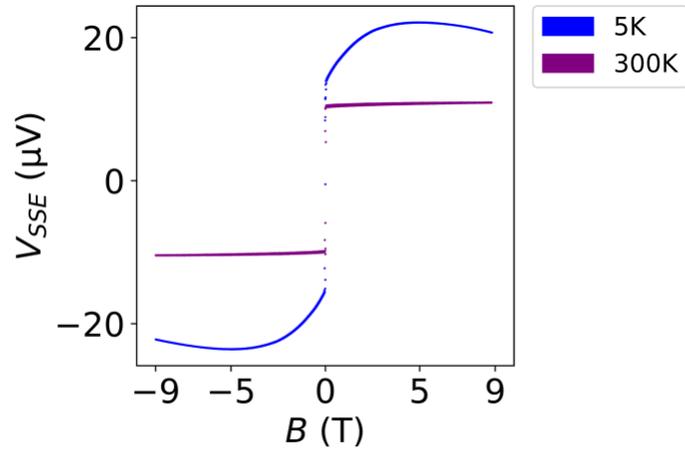

**Figure S2: SSE of the 2 μm wide Rice-made device with field transverse to the wire**
Recorded with a power of 5 mW with the device transverse to the field.

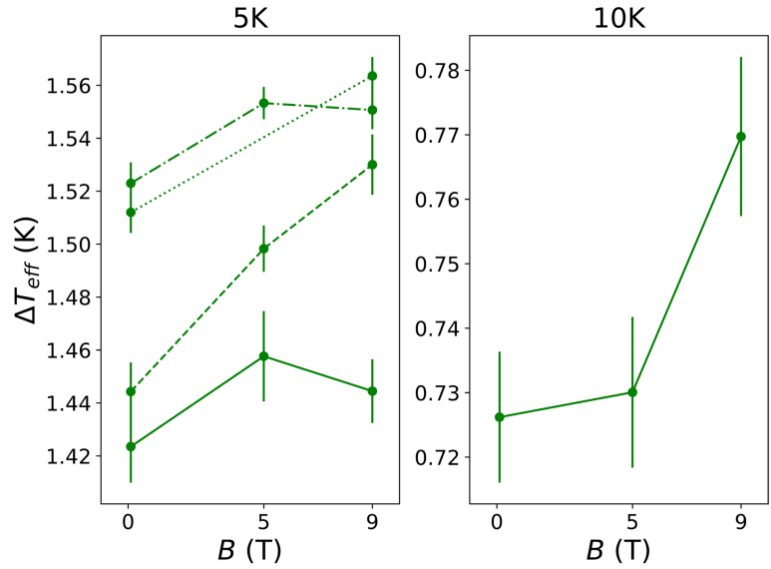

**Figure S3: Additional data on Rice 2 µm wide device**

Recorded with power change from 0 mW to 5 mW. All data were collected with field in-plane in the transverse orientation; no data were collected on this device in the parallel field orientation. Points connected by the same styled line indicate a single run (series of fields at fixed temperature).

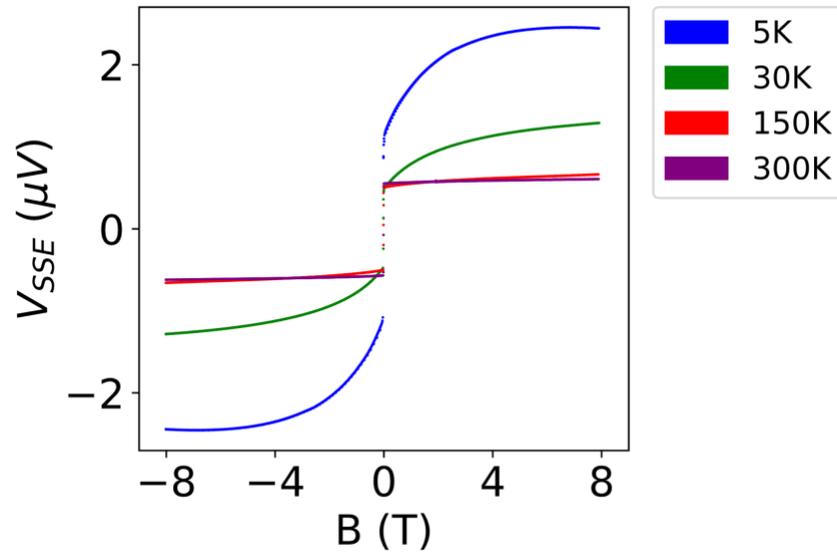

**Figure S4: Spin Seebeck effect on the Argonne-made device**

The data are quantitatively similar to the Rice-made devices, the only difference being the overall magnitude is a factor of 10 weaker. This could be due to many fabrication differences affecting, for example, the spin transfer conductance at the interface between the YIG and the Pt. Recorded with a power of 5 mW with the device transverse to the field.

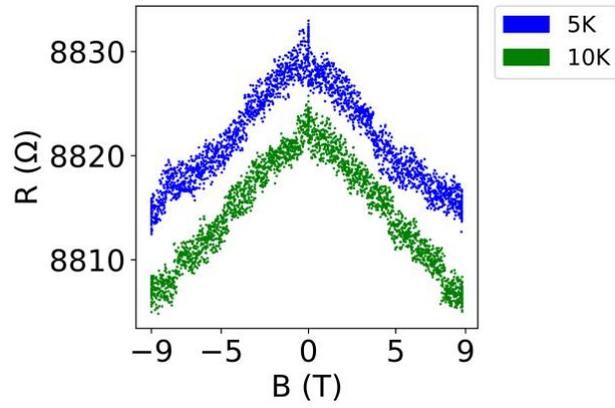

**Figure S5: Magnetoresistance of the Pt wire on the 10 μm wide Rice-made device**
Recorded in the (in-plane) field transverse to wire orientation. The data in the parallel orientation are quantitatively similar though with constant in field offsets due to thermal cycling of the device and changing of contacts.

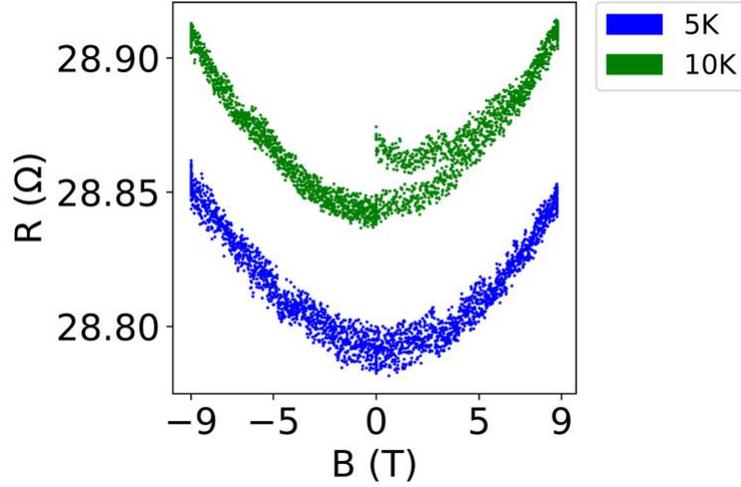

**Figure S6: Magnetoresistance of the Au wire on the 10um wide Rice-made device**
Note that the 10 K data was not totally thermally settled at the beginning of the field sweep run.

We can estimate how much the heater power changes with field due to the magnetoresistance of the Au wire. The overall load consists of the Au wire and two 100 Ω current-limiting resistors $R_{lim}$ in series. In practice we set the constant DC voltage on the overall load to let the heater power at 0 T, $P_0$, to be 5 mW. Now due to the change of the Au heater wire with field, the heater power changes with field as:

$$P(B)/P_0 = \frac{R(B)}{(R(B) + 2R_{lim})^2} \Big/ \frac{R_0}{(R_0 + 2R_{lim})^2}$$

where $R_0$ is the zero-field resistance of the Au heater wire.

When $R(B)$ changes from 28.79 Ω at 0.1 T to 28.84 Ω at 8 T, the heater power increases 0.13%. However, $\Delta T_{eff}$ changes from 2.25 K at 0.1 T to 2.6 K at 8 T, much larger than the heater power change.

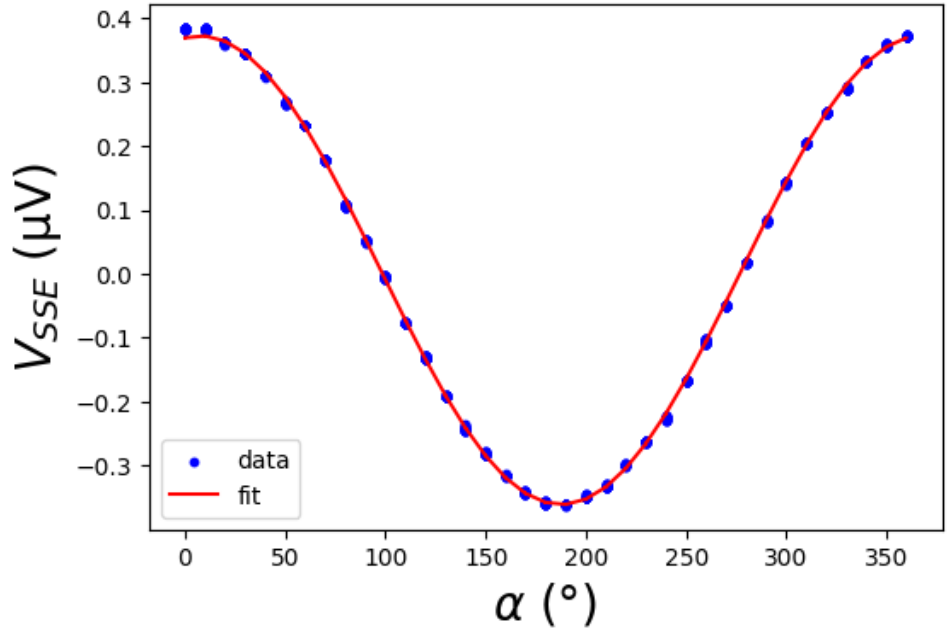

**Figure S7: Spin Seebeck voltage of the 10 μm wide Rice-made device vs. B field angle out of plane**

The angle $\alpha$ is the angle of $\boldsymbol{B}$ from the y axis in the xy plane as defined in Fig. 1b. As expected for the SSE, the angle dependence of $V_{SSE}$ is sinusoidal. The formula for the fit is $V_{SSE} = (a \cos(\beta + b) + c))$ μV where $a = 0.367$, $b = -0.133$, $c = 0.00588$, and $\beta$ should be given in radians. The nonzero $b$ is due to a slight misalignment of the mounted sample relative to the cardinal direction. Measured at 3 K, 2 T, and under a heater power of 0.1 mW.

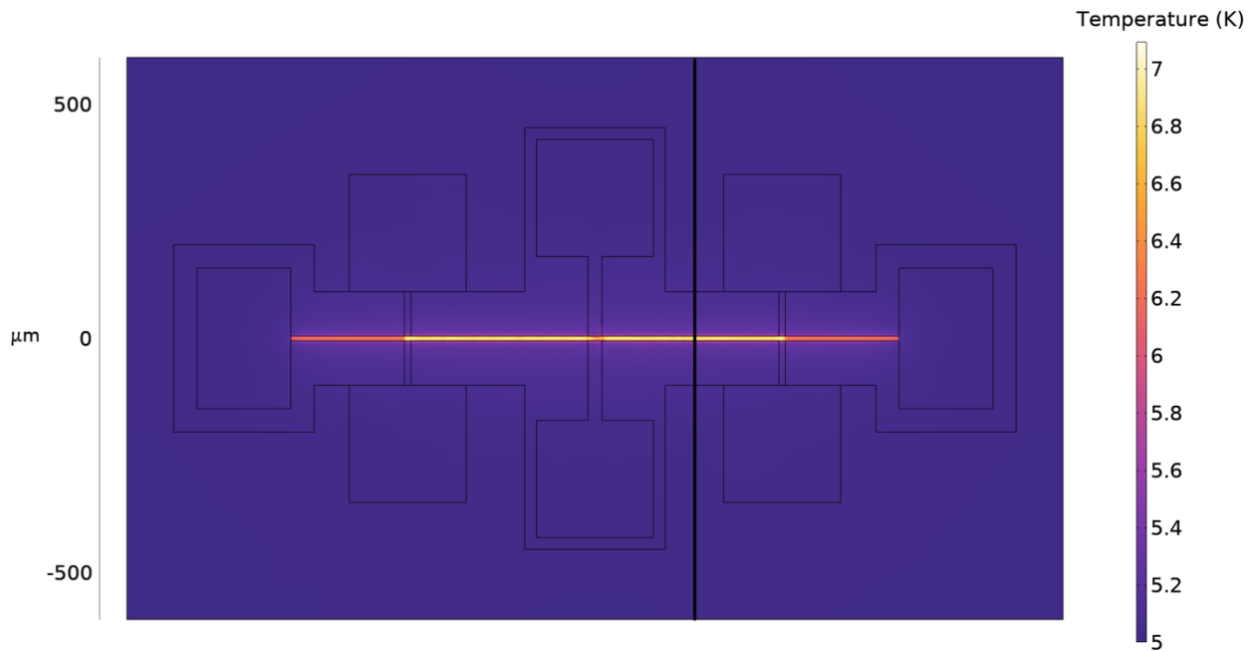

**Figure S8: Top view of thermal model**

A to-scale thermal model of a 6 pad version of the device in COMSOL Multiphysics. In this version the Au wire extends past the Pt wire unlike in Fig. 1a, this, and the addition of pads, don't affect conclusions we make using the model. Here the temperature of the top of the device is shown. The vertical black line is a marker for SM Fig 9.

The electron contribution to the thermal conductivity of the Au (6.56 W m/K) and Pt (0.185 W m/K) were obtained from resistance measurements and the Wiedemann-Franz law, this is taken to be the full thermal conductivity because the measurement regime of ~10 K is well below the Debye temperatures of Au and Pt. The Ti adhesion layer is neglected. The low temperature $SiO_2$ [36], YIG [37], and GGG [44] thermal conductivities used are 0.1, 1, and 200, in units of W/(m K), respectively.

In our low temperature measurement regime the majority (> 90%) of the thermal resistance from the top of the heater wire to the bottom of the substrate actually comes from thermal boundary resistances (BRs). The temperature of the heater is higher over the Pt wire in the model geometry because there the heat (which travels almost entirely downward and not outward, see the arguments of SM Fig. 9) must travel over three metal/dielectric interfaces compared to one.

We roughly model the $Au/SiO_2$, the $SiO_2/Pt$, and the Pt/YIG boundaries (all the metal/dielectric boundaries, that is) as having the same thermal BR, and the Au/Ti thermal BR to be negligible. Since we can't measure these thermal BRs directly, we instead vary the thermal BR in the model until the temperature of the of the Pt detector in the model agrees with the value directly measured using Johnson noise thermometry. For example, for a field of 0 T, substrate/cryostat temperature of 5 K, and heater power of 5 mW we get a Pt temperature of roughly 7 K (as in those conditions in Fig. 3) for a thermal boundary resistivity of $1.5 \times 10^{-6}$ K $m^2/W$.

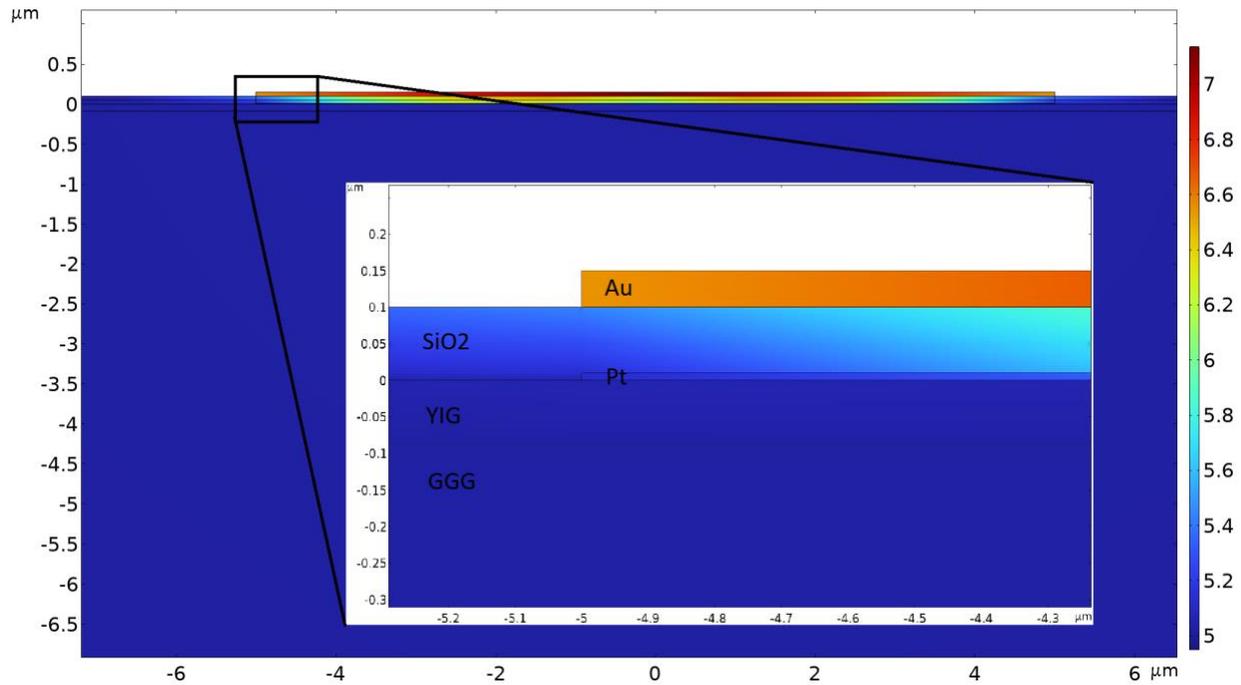

**Figure S9: Cross section of thermal model**

Temperature profile of the device in the *yz* plane at the vertical black line in SM Fig. 8.

The purpose of the thermal model is to determine the dominant thermal paths in the system, and to evaluate three principal concerns: (1) that there is significant unknown heat bleeding out sideways through the SiO$_2$ insulator before the heat current reaches the GGG directly below the Pt. This would be an issue for an absolute, quantitative measurement of the YIG spin Seebeck coefficient (though it is irrelevant to the main results of our analysis, that the observed field dependence of the excess noise is not compatible with spin shot noise and that such spin shot noise is likely not measurable with this approach). Side-ways heat flow can be estimated from the model, and, for the length of heater over the Pt wire, roughly 86% of heater power ultimately enters the YIG vertically through the Pt). (2) That some heat entering the Pt through the top, escapes through its sides, leading in-plane temperature gradients in the Pt. This would be a problem if we were measuring the Pt temperature gradient to calculate say the Nernst-Ettingshausen effect, however from the model we find this percentage to be very small, order of 0.88%. Finally, (3) that there might be significant lateral (in plane) heat flow in the metals which could allow the bonding pads to sink heat, also causing overestimation of the thermal gradient in the substrate.

Addressing the final concern: consider a 10 μm wide device. From under the heater to a pad, the Pt wire covers a length of roughly 100 μm with a width of 10 μm and a thickness of 7.5 nm, using $\rho_{Pt,thermal} = 5.41$ m K/W gives a total thermal resistance from the heated portion of the Pt to the pad of $7.21 \times 10^9$ K/W. For downwards heat travel we take the thermal boundary resistivity of $1.5 \times 10^{-6}$ K m$^2$/W from the model and divide it by the area of contact between the Pt wire and the YIG film, roughly 1000 µm x 10 µm giving a total thermal resistance of $10^3$ K/W. Clearly it is far easier for the heat to travel downwards than laterally. The Pt wire (and the Au wire, following the same argument) is just far too thin for lateral heat transport to be significant. This demonstrates no significant heat escapes through the wirebonds to the pads.

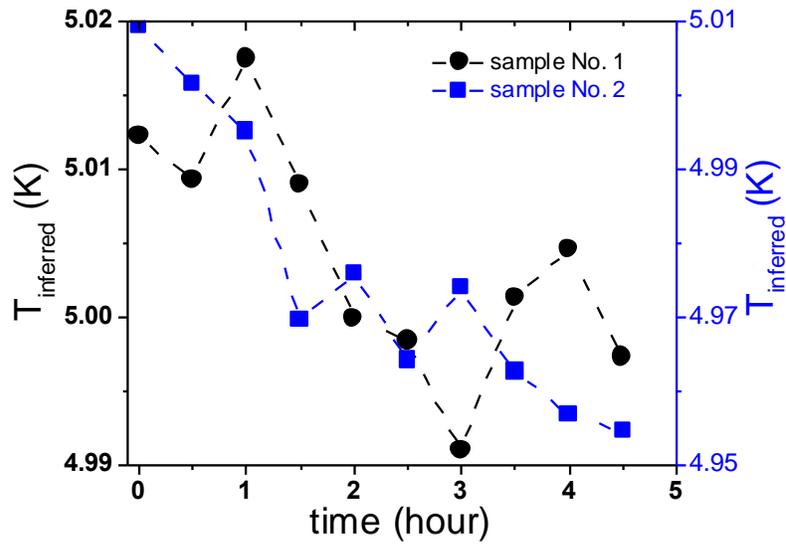

**Figure S10: Temperature stability**

By performing sequential Johnson-Nyquist noise measurements in the absence of any heater power, it is possible to assess the temperature stability of the noise probe and sample. The data above demonstrate that after waiting for an hour for conditions to stabilize, the temperature stability is within a few mK per hour at a cryostat temperature of 5 K. The protocol employed to obtain the data in Fig. 2 of the main text involved repeated measurements at zero heater power and target heater power, to mitigate any long-term drift.

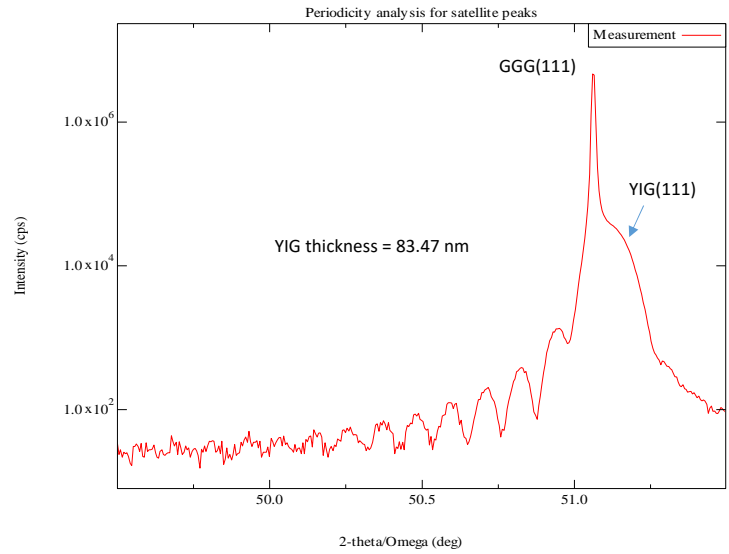

**Figure S11: Film quality**

Diffraction of the YIG film on GGG substrate. The film is smooth and of high quality, with a confirmed thickness of 83.47 nm.